\def\articleTitle{Latent Variable Models for Distributional Features}

    \documentclass[letterpaper,12pt]{article}
    \usepackage{abstract}

    \usepackage{titlesec} 
    \usepackage{authblk} 
    \usepackage{datetime}
    \newdateformat{usvardate}{\monthname[\THEMONTH] \ordinal{DAY}, \THEYEAR}
  	\usepackage[bottom]{footmisc}
    \usepackage{fancyhdr}
    \fancyheadoffset{0cm}
    \providecommand{\keywords}[1]{\textbf{\textit{Keywords---}} #1}
\newcommand\wordcount{%
  \immediate\write18{texcount -utf8 -merge -sum -incbib -dir -sub=none -brief \jobname.tex | cut -d : -f 1 > 'count.txt'}%
  \input{count.txt}\ignorespaces words%
}
    \usepackage{amssymb}
    \usepackage{amsthm}
    \usepackage{newtxmath}
    \usepackage{bm}
    \usepackage{upgreek}
    \newcommand{\Btheta}{\boldsymbol{\uptheta}}
    \newcommand{\Balpha}{\boldsymbol{\upalpha}}
    \newcommand{\Bbeta}{\boldsymbol{\upbeta}}

    \usepackage{microtype}
    \usepackage[utf8]{inputenc}
    \usepackage{newtxtext}
    \usepackage{setspace}
    \usepackage{lineno,xcolor}
	\usepackage[top=1.5in, bottom=1.5in, left=1in, right=1in]{geometry}

	\usepackage[colorinlistoftodos]{todonotes}
	\usepackage{soul}
    \usepackage{graphicx}
    \usepackage{float}
    \usepackage{tikz}
    \usepackage{tikz-cd}
    \usepackage{comment}
    \usetikzlibrary{positioning}
    \usetikzlibrary{bayesnet}
    \usetikzlibrary{arrows}
    \usepackage{printlen}
    \usepackage{booktabs}  
    \usepackage{longtable}
    
    \usepackage{tabularx}
    \newcolumntype{L}[1]{>{\raggedright\arraybackslash}p{#1}}
    \newcolumntype{C}[1]{>{\centering\arraybackslash}p{#1}}
    \newcolumntype{R}[1]{>{\raggedleft\arraybackslash}p{#1}}
    \usepackage{relsize}
    \usepackage{tabularray}
    \usepackage{multirow}
    \usepackage{hyperref}
    \usepackage{csquotes}
    \usepackage[style=apa6,backend=biber]{biblatex}
    \bibliography{_bdlvm-article02}
    
\begin{document}

  \title{\vspace{-15mm}\fontsize{21pt}{10pt}\selectfont\textbf{\articleTitle}}

  \author{Luna Fazio$^{1*}$, Paul-Christian Bürkner$^{1}$}
  \renewcommand\Authands{ and }

\date{\small
  $^1$ Department of Statistics, TU Dortmund University, Germany\break
  $^*$ Corresponding author; Email: \texttt{bmfaziol@gmail.com}
}
\maketitle

\begin{abstract}
\noindent Analyzing the mean response of study subjects in psychological research is a standard, well-justified practice. However, theoretical arguments and empirical evidence also suggest that there is value in investigating other aspects of the distribution of such responses, such as their variability or skewness.

A particular challenge that practitioners face is statistical modeling of associations between distributional features and other outcomes of interest. The most common approach is to perform estimation in two steps: distributional features are estimated first, and then those estimates are used as predictors for the relevant outcomes. Such an approach is most amenable to implementation in standard statistical software, but it ignores estimation error and can therefore lead to biased estimates and increased error rates.

We introduce Distributional Feature Latent Variable Models (DFLVM), a general framework that represents between-person difference in distributional features as random intercepts. These intercepts can be simultaneously used as predictors for downstream outcomes and their associations estimated in a single estimation step. We compare the performance of our approach against two-step procedures in a simulation study and through a re-analysis of a real dataset.

\end{abstract}

\setlength\parindent{0in} \keywords{Psychometrics, distributional regression, structural equation modeling, Bayesian inference}


\section{Introduction}
\label{sec:introduction}

A common approach to measurement in psychology consists of taking the mean or sum score of participant responses on some multi-item scale \parencite{sijtsmaRecognizeValueSum2024}. Participants that attain the same score may simultaneously exhibit very different item-level response patterns, and those differences can be characterized through various statistical summaries such as the standard deviation, skewness, or range of the responses. We will generically refer to such features of the data which go ``beyond the mean" as \emph{distributional features}. A natural question is whether inter-personal differences in distributional features reflect valuable information for psychometric measurement.

This question has been approached from two distinct traditions. In the psychometric literature, recent empirical investigations have looked into the added predictive value of distributional features computed directly from scale responses, with mixed results \parencite{maderEmotionalInstabilityNeuroticism2023,nielsenMeanCanWe2023,dejonckheereComplexAffectDynamics2019,houbenRelationShorttermEmotion2015}. Meanwhile, in the cognitive modeling literature, behavioral data is analyzed with process models whose parameters relate substantively meaningful parameters \parencite[such as drift rate and boundary separation in the drift-diffusion model;][]{ratcliffDiffusionDecisionModel2008} that are simultaneously related to the mean and other distributional features of the response. The resulting estimates can then be used as predictors of downstream outcomes such as intelligence \parencite{frischkornCognitiveModelsIntelligence2018}. Both traditions share the underlying idea that richer characterizations of individual response distributions can carry meaningful person-level information that is not captured by the mean alone.

Our paper is motivated by a methodological limitation that has been recognized across both traditions. When distributional features -- whether summary statistics of scale responses or estimated parameters of a cognitive model -- are treated as directly observable quantities, this entails a two-step estimation process in which a point estimate of the distributional feature is first obtained and then used as a predictor or covariate. Because such an approach ignores estimation error, it does not adequately propagate uncertainty, leading to inflated Type I error and negative bias on the associated parameters \parencite{jahngAnalysisAffectiveInstability2008,frischkornWorstPerformanceRule2016}. \textcite{frischkornCognitiveModelsIntelligence2018} explicitly identify this problem in the cognitive modeling tradition and recommend hierarchical Bayesian estimation as the principled remedy, while noting that practical adoption remains limited by the scarcity of accessible software implementations.
This problem has been similarly recognized in the psychometric literature: \textcite{wangInvestigatingInterindividualDifferences2012} investigate the association between inter-individual variability in negative affect and health outcomes, and note the aforementioned issues with using a two-step procedure to obtain a point estimate of variability before regressing it on another outcome. While the authors in this case also implement a Bayesian single-step procedure, their implementation is specific to the analysis at hand and would demand additional expertise from researchers wishing to build on it. For instance, a key result from \textcite{maderEmotionalInstabilityNeuroticism2023} is that failing to account for floor and ceiling effects in item scale responses leads to losses in power, but dealing with this requires the specification of a censored observational model.

We propose a general modeling framework where distributional features are treated as latent variables, represented via person-specific random intercepts. This enables single-step estimation that fully propagates uncertainty. Further, it can accommodate more sophisticated modeling choices (e.g., censoring), allowing it to be readily applied to both item response data in a psychometric context as well as behavioral measurements used in cognitive process models.

\subsection{Distributional features in psychology}
\label{sec:introduction-roles}

To situate our contribution more broadly, we briefly review the two main roles that distributional features have played in psychological research and give examples of each.

There are two ways in which distributional features can be used in an analysis which we wish to distinguish for clarity.  On the one hand, distributional features can be of direct interest as part of the model for a response variable. A classic example is that of a heteroscedastic linear regression model, in which the scale parameter of the response distribution can be predicted separately from the mean parameter \parencite{harveyEstimatingRegressionModels1976}. More generally, such models fall under the framework of generalized additive models for location, scale and shape (GAMLSS, \cite{stasinopoulosFlexibleRegressionSmoothing2017}), also referred to as distributional regression models (\cite{fahrmeirRegressionModelsMethods2021}, Chapter 10; \cite{burknerBayesianItemResponse2021}). On the other hand, distributional features can instead be used as explanatory variables that predict other outcomes in a model. This is a less common approach for which there does not appear to be a general naming convention, so we adopt the term \emph{distributional covariates} to refer to such an approach. In any case, both approaches have applications in psychology and we review some examples below.

For starters, the classic definition of measurement invariance given by \textcite{meredithMeasurementInvarianceFactor1993} contemplated the possibility of response models with general parametric differences between subpopulations. A statistical model to account for those differences was introduced in \textcite{bauerPsychometricApproachesDeveloping2009} (Moderated Nonlinear Factor Analysis, shortened as MNLFA), which can be understood as involving a distributional regression component on the variance parameters of a factor model. More recently, \textcite{wiedermannDistributionalCausalEffects2022} demonstrated that distributional regression can be used for detecting effects in interventional settings in addition to those that appear on the mean, or even when the latter are absent. Clearly, the distributional regression perspective is one with a long-standing presence in the psychological literature and one for which well-established estimation frameworks are widely available.

In contrast, when it comes to the use of distributional covariates, the literature does not appear to have coalesced around a common modeling framework. A long line of research has investigated whether the variability of responses, both across items within a single scale administration and across repeated measurements over time, carries predictive value for outcomes such as well-being and adjustment \parencite{bairdNatureIntraindividualPersonality2006,wangInvestigatingInterindividualDifferences2012,houbenRelationShorttermEmotion2015}, with more recent work examining this question in a systematic and broad-ranging manner \parencite{dejonckheereComplexAffectDynamics2019, nielsenMeanCanWe2023}. The general finding has been that few among the many possible distributional summaries add meaningful predictive value beyond the mean. This inconclusive picture may partially be an artifact of i) inappropriate handling of estimation uncertainty leading to inflated error rates and biased estimates \parencite{jahngAnalysisAffectiveInstability2008} and ii) loss of power due to omission of floor and ceiling effects \parencite{maderEmotionalInstabilityNeuroticism2023}. The lack of freely available, user-friendly, and sufficiently flexible frameworks for modeling of distributional covariates is a non-trivial obstacle for practitioners, which could partly explain why such issues persist even in recent literature.

\subsection{Our contributions}

\begin{enumerate}
\item We propose an estimation framework that represents between-person variation in distributional features as latent variables and simultaneously allows their use as downstream covariates.
\item We investigate the impact of using these distributional covariates in single-step estimation compared with two-step estimation approaches that only use point estimates of distributional features.
\item We evaluate our framework both via extensive simulation studies and analyses of real-world data.
\end{enumerate}

\section{Method}

Previous work investigating distributional features has largely focused on direct summaries of the data such as observed variances, skewness, response ranges, among other more sophisticated metrics (see \cite{dejonckheereComplexAffectDynamics2019,nielsenMeanCanWe2023}). We take a different approach, and propose the use of latent variable models to represent the information contained in distributional features, a framework which we refer to as a \emph{Distributional Feature Latent Variable Model} (DFLVM). This framing facilitates a general discussion of the statistical issues at play during estimation, and allows for more direct connections to be drawn between the theories that motivate the use of distributional features and their formal implementations.

In this section, we formally introduce the DFLVM framework and provide an overview of the statistical considerations that their use raises. We defer a discussion of substantive implications of using latent variable models for distributional features to Section~\ref{sec:discussion}.

\subsection{Measurement model for distributional features}

Consider a setting with $N$ participants indexed by $n\in\{1,\dots,N\}$, each of which has been measured a total of $M_n$ times through some observable $X$. Let the probability distribution of $X$ be denoted by $\text{D}(\Btheta)$, parameterized in terms of a vector of $P$ parameters $\Btheta = (\theta_1, \dots,\theta_P)$. We use a hierarchical structure to capture person- and parameter-specific variation. Let $\mathbf{U} = (U_{11},\dots,U_{NP})$ be a vector of normally distributed random intercepts with length $N\times P$ and covariance matrix $\Sigma$. Then, observations of $X$ can be used to construct a the measurement model for a distributional feature:
\begin{equation}
\label{eq:modelx}
\begin{aligned}
\mathbf{U} &\sim \text{Normal}(\mathbf{0}, \Sigma)\\
\theta_{np} &= g_p^{-1}(\alpha_p + U_{np})\\
X_{nm} &\sim \text{D}(\Btheta_{n})
\end{aligned}
\quad
\begin{aligned}
\\
n &\in \{1,\dots,N\}, \; p \in \{1,\dots,P\} \\
m &\in \{1,\dots,M_n\},
\end{aligned}
\end{equation}
where $\alpha_p$ is the global intercept of its corresponding parameter and $g_p$ is some appropriate link function. The components of the vector $\mathbf U$ will be later used as distributional covariates.

We do not require the measurements $X_{nm}$ to be univariate. For example, evidence accumulation models \parencite{ratcliffDiffusionDecisionModel2008} take bivariate measurements (reaction time + decision) and often have multiple distributional parameters, for each of which a measurement model can be constructed.

From a statistical perspective, the distributional covariates $\mathbf U$ simply are random intercepts that appear on the linear predictors of parameters $\Btheta$, but this also corresponds to a latent variable representation. To see this, consider the case where $P=1$. Frequently, the first parameter of a distribution corresponds to its mean, in this case represented by $\theta_1$. Then, the model in Equation \ref{eq:modelx} is equivalent to a simple confirmatory factor analysis (CFA) model, where $U_{n1}$ is the latent trait for subject $n$, which loads on the set of measurements $X_{nm}$ with factor loadings fixed to 1. That is a well-studied setting, so our interest in developing DFLVMs lies in the case where $P>1$, with parameters that are related to the distributional features of the measurements.

Given the comparison drawn above, a natural question is whether factor loadings can also be estimated in a DFLVM. One can readily show that this is possible for the scale parameter of a normally distributed response (see Section~2.2 of \cite{fazioGaussianDistributionalStructural2025}), but this is not the focus of the present paper.

\subsection{Distributional features as distributional covariates}
\label{sec:ddm-covariates}

As detailed in Section~\ref{sec:introduction-roles}, we are interested in the case where variation in distributional features can be used to predict other outcomes. For the purposes of this paper, a generalized linear regression model suffices to demonstrate the relevant principles of this second modeling step.

Building on the setup for Equation~\ref{eq:modelx}, consider a random variable $Y$ with distribution $\text{D}'$, a mean parameter $\mu$, a link function $g'$, and additional parameters $\mathbf{\xi}$, which has been collected from the same set of $N$ participants. Then, the expected value of $Y$ is related to the distributional features of $X$ through
\begin{equation}
\begin{aligned}
\label{eq:modely}
Y_n &\sim \text{D}'(\mu_n,\, \mathbf{\xi})\\
\mu_n &= g'^{-1} (\beta_0 + \beta_1 U_{n1} + \dots + \beta_P U_{nP}).
\end{aligned}
\end{equation}
We refer to this component as the second-stage model. Taken together, Equation~\ref{eq:modelx} and Equation~\ref{eq:modely} form the DFLVM. We are primarily interested in estimating the regression coefficients $\beta_{\{1,\dots,P\}}$ that quantify the impact of the distributional covariates $\mathbf{U}$ on the second-stage outcome $Y$. The notation above is kept simple for expository purposes. The outcome $Y$ can be either a manifest or latent variable (in which case we leave its measurement model unspecified here) and it may admit additional parameters $\mathbf{\xi}$ other than the mean $\mu$ on which distributional covariates could be placed.

\subsubsection*{Standardized coefficients}

In some circumstances (e.g., when investigating within-person associations over time, \textcite{wangStandardizingWithinPersonEffects2019}), it may be of interest to compare a standardized version of the coefficients $\beta_{\{1,\dots,P\}}$ instead of their raw values. Standardization can be conducted relative to the scale of the covariates, the response, or both. The methods we compare in this paper only vary in how the estimate and process the covariates $\mathbf{U}$, so we standardize by the scale of the covariates only:
\begin{align}
\label{eq:stdcoeff}
b_p = \beta_p\times\text{SD}(U_p),
\end{align}
where $b_p$ represents the standardized version of $\beta_p$ and $\text{SD}(U_p)$ is the standard deviation of the corresponding distributional covariate. If written in terms of estimates, rather than true values, the estimates $\hat{b}_p$ of the standardized coefficients $b_p$ are computed as
\begin{align}
\label{eq:stdcoeff-est}
\hat{b}_p = \hat{\beta}_p \times \widehat{\text{SD}(U_p)},
\end{align}
where both $\hat{\beta}_p$ and $\widehat{\text{SD}(U_p)}$ are specific to the estimation method (see below).

\subsubsection*{Form of the distributional covariate}

When using distributional features as covariates, the analyst must make a choice regarding the form in which its value enters into the predictor. An approach commonly taken in the literature is to take data summaries as predictors. Because of the many possible summaries that one can concoct, this can introduce a large number of researcher degrees of freedom \parencite{simmonsFalsePositivePsychologyUndisclosed2011}. For instance, Table~1 of \textcite{dejonckheereComplexAffectDynamics2019} illustrates nine different summaries that have been proposed for the analysis of affect dynamics. In the case of \textcite{nielsenMeanCanWe2023}, eleven summaries beyond the mean are considered, including the standard deviation, range, and skew. By contrast, \textcite{wangInvestigatingInterindividualDifferences2012} takes a model-based approach and uses the random intercepts that represent inter-personal differences in the distributional parameter of interest. The differing approaches illustrate that the choice of form in which to represent the distributional covariate is not obvious. Our proposed DFLVM formulation more explicitly frames that choice as part of the model-building exercise by introducing link functions which relate distributional parameters to person-specific latent variables. In the case of, e.g. standard deviations, trying a log link would be a natural first choice, while a scaled logit link could be suitable for correlations. That said, constructing a parametric representation of general data summaries is not always a straightforward task, but we defer discussion of this point to Section~\ref{sec:discussion}. Our results in Section~\ref{sec:sims} demonstrate that choosing to use a distributional feature directly in place of the latent person-specific component can lead to very different assessments of model fit.

\subsection{Estimation}
\label{sec:estimation}

In this paper, all models are estimated within the Bayesian framework. In a Bayesian estimation procedure, one aims to obtain a posterior probability distribution over plausible parameter values given the observed data. This requires defining both a likelihood function and a prior distribution (for an introduction to Bayesian inference, see \cite{mcelreathStatisticalRethinkingBayesian2020}). Using $\pi$ to denote a probability density function, we can write the joint likelihood implied by Equations~\ref{eq:modelx} and \ref{eq:modely} as
\begin{align}
\label{eq:likelihood}
\begin{split}
L(\mathbf{x}, \mathbf{y} &\mid \mathbf{U}, \Sigma, \Balpha, \Bbeta, \mathbf{\xi}) = \\
&\left[
    \prod_{n=1}^N
    \pi(y_n\mid \Bbeta, \mathbf{U}, \mathbf{\xi})
    \left(
        \prod_{m=1}^M\pi(x_{nm}\mid\Btheta_n)
    \right)
    \left(\prod_{p=1}^P
        \pi(\theta_{np} \mid \alpha_p,\,U_{np})
    \right)
\right]
\pi(\mathbf{U}\mid\Sigma).
\end{split}
\end{align}
Leaving the priors unspecified for now, we can write the posterior up to proportionality as
\begin{align}
\label{eq:posterior}
\pi(\mathbf{U}, \Sigma, \Balpha, \Bbeta, \xi \mid \mathbf{x}, \mathbf{y}) \propto
L(\mathbf{x}, \mathbf{y} &\mid \mathbf{U}, \Sigma, \Balpha, \Bbeta, \xi)\pi(\Sigma)\pi(\Balpha)\pi(\Bbeta)\pi(\xi).
\end{align}
The main practical challenge for Bayesian inference is computation of the posterior distribution, which generally requires the use of approximation methods. We use the Markov chain Monte Carlo (MCMC) sampler implemented in the Stan probabilistic programming language \parencite{standevelopmentteamStanReferenceManual2026}. The output of MCMC is a sequence of samples (known as a \emph{chain}) intended to represent the target distribution, but the quality of the approximation always needs to be verified. We describe the criteria we use to assess convergence at the end of Section~\ref{sec:perfmetrics}.

\subsubsection*{Prior specification}

In the ideal Bayesian workflow, all model parameters are given priors that represent a relevant state of knowledge that will be updated through the likelihood as new data arrives \parencite{gelmanBayesianWorkflow2020,koenigEditorialMovingNoninformative2021}. The purpose of this paper, however, is to investigate general conditions for which estimation of a DFLVM is viable. Hence, we use very weakly informative priors in our case studies \parencite{gelmanPriorCanGenerally2017}. We show the exact priors being used along with the model descriptions in Section \ref{sec:sims} and \ref{sec:casestudy}.


\subsection{Two-step approaches and estimation error}

As discussed in Section~\ref{sec:introduction-roles}, two-step estimation is the most commonly used approach for models that involve distributional covariates, even if this is not always explicitly framed as such. For instance, the various data summaries investigated in \textcite{nielsenMeanCanWe2023} can be seen as the result of a first stage where a point value for the distributional feature is estimated, followed by a second stage where those estimates are used as distributional covariates, as if they had been directly observed. Another approach, discussed by \textcite{wangInvestigatingInterindividualDifferences2012}, proceeds by estimating the random intercepts that represent the person-specific component of the observed variability in negative affect. Those point estimates are then used as distributional covariates to predict health outcomes. The authors recognized that the approach does not propagate the uncertainty in the random intercepts, and additionally developed a model for single-step Bayesian estimation. The advantage of two-step approaches for practitioners is that each stage can generally be reduced to a model simple enough to fit within already-existing estimation frameworks. This is illustrated by the more recent example of \textcite{maderEmotionalInstabilityNeuroticism2023}, where the authors also investigated between-person variability in negative affect, this time relating it to neuroticism. While they do not cite \textcite{wangInvestigatingInterindividualDifferences2012}, they converged to essentially the same two-step procedure, both acknowledging the issue of uncertainty propagation and that a single-step approach would require a custom implementation that goes beyond the types of model that can be represented in standard software.

To elaborate on the matter of uncertainty propagation, the issue stems from the fact that distributional features have to be estimated from data, and hence are subject to estimation error. A central finding from the measurement error literature is that using covariates measured with error in a regression model without accounting for that can lead to biased and inconsistent estimates of the corresponding coefficient, and to inflated test error rates. \textcite{gustafsonImpactUnacknowledgedMeasurement2021} provides a comprehensive overview of key results about the impact of measurement error under various conditions. Estimation error constitutes a particular kind of measurement error in this context, and hence introduces the aforementioned problems. At the same time, because estimation error decreases with the number of repeated measurements, its practical effect may be ignorable in certain circumstances \parencite{ramExaminingInterindividualDifferences2005}. We investigate these issues in Section \ref{sec:sims} via simulations.

\section{Simulations}
\label{sec:sims}

We conduct extensive simulations to assess the performance of single-step DFLVM estimation as compared with two other two-step procedures. One approach, exemplified by \textcite{nielsenMeanCanWe2023}, takes the person-specific value of a distributional feature (in this case a standard deviation) directly as the distributional covariate. The other approach, demonstrated by \textcite{wangInvestigatingInterindividualDifferences2012} and \textcite{maderEmotionalInstabilityNeuroticism2023}, estimates random intercepts that model the between-person variation in the distributional feature and then uses their point estimates as the distributional covariate. For brevity, we will henceforth refer to the first approach as Full Parameter (FP) and to the second one as Latent Component (LC).

\subsection{Computational setup}
\label{sec:compsetup}

Our simulations were implemented in the \texttt{R} programming language \parencite{rcoreteamLanguageEnvironmentStatistical2023}. We specified our models in the \texttt{brms} package \parencite{burknerBrmsPackageBayesian2017}, which provides a user-friendly interface for generation of Stan code \parencite{standevelopmentteamStanReferenceManual2026}.\footnote{The functionality required to specify a DFLVM-type model is not available on an official release at the time of writing. The development version used in the present work can be accessed at \url{https://github.com/paul-buerkner/brms/tree/brms3}} To facilitate reproducibility, the simulation pipeline is implemented using the \texttt{targets} framework \parencite{landauTargetsPackageDynamic2021}. Full code is available via an online repository.\footnote{See the folder \texttt{simulation-study} at \url{https://github.com/bdlvm-project/dflvm-paper}}
\subsection{Performance metrics}
\label{sec:perfmetrics}

The coefficients $\Bbeta = \beta_{\{1, \dots,P\}}$, which express the impact of the distributional covariates on the outcome $Y$, are the primary parameters of interest in contexts where a DFLVM would be applied. Therefore, we focus on assessing estimation performance for those parameters. We consider both the unstandardized coefficient $\beta$ and its standardized version $b$, as defined in Equation~\ref{eq:stdcoeff}. In the case of the DFLVM, standardization is performed during fitting on a per-sample basis, whereas two-step models use the standard deviation of the vector of point estimates of the distributional covariates, which is obtained during the first stage.

To evaluate parameter recovery, we use bias and the Root Mean Squared Error (RMSE). If $S$ is the number of simulations, $\hat \beta^{(s)}$ the estimated posterior mean, and $\beta^{(s)}$ the true coefficient of the $s$th simulation, we have
\begin{align}
\text{Bias} = \frac{1}{S} \sum_{s=1}^S (\hat\beta^{(s)} - \beta^{(s)}),\\
\text{RMSE} = \sqrt{\frac{1}{S} \sum_{s=1}^S(\hat\beta^{(s)} - \beta^{(s)})^2}.
\end{align}

To evaluate uncertainty calibration, we first investigate the empirical type I error rate and power from the posterior credible intervals for a nominal type I error rate of $5\%$. Additionally, to obtain a holistic measure of uncertainty calibration, independent of any nominal error rate, we apply the approach of \textcite{sailynojaGraphicalTestDiscrete2022}. Motivated by simulation-based calibration, this approach considers the rank of the true value within the posterior distribution across all the posteriors estimated in the simulation. It then estimates the probability of observing the most extreme point on the empirical cumulative distribution function of those ranks compared to a discrete uniform distribution of the same size. In addition to a graphical representation, it emits a statistic for formal testing of calibration named $\log \gamma$. For mathematical details, see \textcite{sailynojaGraphicalTestDiscrete2022}.

%
%

Following one of the evaluations performed in \textcite{nielsenMeanCanWe2023}, we further investigate out-of-sample predictive performance using the coefficient of determination evaluated on test data. Using the notation in Equation~\ref{eq:modely}, the test data is a vector of new responses $Y^*_{\{1,\dots,_N\}}$ generated from the same distribution $\text{D}'$ with the distributional covariates $U_{\{1,\dots,N\}}$ held fixed. For each simulation $s\in\{1,\dots,S\}$, we calculate the ratio of the variance of the vector of estimated posterior means $\hat \mu^{(s)}_{\{1,\dots,N\}}$ to the variance of the vector of corresponding new responses $Y^{*(s)}_{\{1,\dots,N\}}$:
\begin{equation}
R^2_s = \text{Var}\left(\hat\mu^{(s)}_{\{1,\dots,N\}}\right)/ \, \text{Var}\left(Y^{*(s)}_{\{1,\dots,N\}}\right).
\end{equation}

Finally, fitting a more complex model such as the DFLVM can be computationally demanding, so we compare the wall time of the first-stage fit of the two-step LC approach against the single-step DFLVM fit. Because the second-stage fit of the LC approach reduces to a simple generalized linear model in our case studies, it only adds a negligible amount of time to the fitting process.

We only calculate performance metrics over those fits where convergence was achieved as assessed via Stan's built-in diagnostics. Specifically, we discard any fits with $\hat R$ above 1.05, divergent transitions above 50 or Effective Sample Size below 400 \parencite{vehtariRankNormalizationFoldingLocalization2021}. More details on these diagnostics can be found in the Stan online user guide \parencite{standevelopmentteamStanReferenceManual2026}.

\subsection{Simulation study: Variability as predictor}

In \cite{maderEmotionalInstabilityNeuroticism2023}, the authors explore the association between neuroticism and mood variability across various longitudinal datasets, and our simulations are inspired by this real-world case. Their main analysis uses a hierarchical distributional regression model with baseline neuroticism as a predictor of the mean and variability of daily mood levels, but they additionally present a model where the mean and variability of mood is used to predict mean neuroticism. They use a two-step estimation procedure, where the first stage estimates person-specific random intercepts for the mean and variability of mood:
\begin{equation}
\begin{aligned}
\label{eq:madermodel1}
\mathbf{R} = \begin{pmatrix}
1 & \rho \\
\rho & 1
\end{pmatrix}, \quad
\mathbf{D} &= \begin{pmatrix}
\tau_{11} & 0 \\
0 & \tau_{22}
\end{pmatrix}, \quad
\Sigma = \mathbf{D \, R \, D}\\
(U_{n1},\, U_{n2})^\intercal &\sim
    \text{Normal}(\mathbf{0},\Sigma)\\
\mu_n = \alpha_1 + U_{n1}, & \quad
\sigma_n = \exp(\alpha_2 + U_{n2})\\
\text{Mood}_{nm} &\sim
    \text{Normal}(\mu_n,\,\sigma_n).
\end{aligned}
\end{equation}
The original analysis then used the posterior means $\overline{U}$ of the random intercepts to predict mean neuroticism in a separately estimated, second-stage model:
\begin{equation}
\label{eq:madermodel2}
\text{Neuroticism}_{n} \sim \text{Normal}(\beta_0 + \beta_1 \overline{U}_{n1}+ \beta_2 \overline{U}_{n2}, \xi)
\end{equation}
In our simulations, we will not only consider a normal likelihood case (with identity link) for the second-stage model, but
also investigate a Poisson likelihood with log link and a Binomial likelihood with logit link, to study the effects of varying likelihood and link functions. We write
\begin{equation}
\label{eq:madermodel2}
\text{Neuroticism}_{n} \sim D'(g(\beta_0 + \beta_1 \overline{U}_{n1}+ \beta_2 \overline{U}_{n2}), \xi),
\end{equation}
where $D'$ represents the chosen likelihood, $g$ the inverse link function, and $\xi$ a possible second likelihood parameter (residual standard deviation in the normal case, fixed number of events in the Binomial case, and unused in the Poisson case).

Independent of likelihood and link choice, we refer to the procedure of propagating only posterior means of random intercepts as the Latent Component (LC) approach. 

In contrast, in our proposed DFLVM approach, we estimate both model stages jointly, directly using the $U$ as predictors in the second-stage model:
\begin{equation}
\label{eq:DFLVM}
\text{Neuroticism}_{n} \sim D'(g'^{-1}(\beta_0 + \beta_1 U_{n1}+ \beta_2 U_{n2}), \xi).
\end{equation}
An additional approach of interest is that of using the value of the mean $\mu$ and variability parameter $\sigma$ by itself, instead of isolating the person-specific components $U$. Accordingly, the second-stage model, again estimated separately, is given by
\begin{equation}
\label{eq:nielsen}
\text{Neuroticism}_{n} \sim D'(g(\beta_0 + \beta_1 \mu_{n} + \beta_2 \sigma_{n}), \xi).
\end{equation}
Since the full value of each parameter is used directly in the model, we refer to this as the FP (Full Parameter) approach.
We also consider an \textit{oracle model}, which uses the true values $U^*$ of the distributional features -- known only for simulated data -- as covariates:
\begin{equation}
\label{eq:oracle}
\text{Neuroticism}_{n} \sim \text{Normal}(\beta_0 + \beta_1 U^*_{n1}+ \beta_2 U^*_{n2}, \xi).
\end{equation}
Since its covariates are known, the oracle only has a second-stage fit. For $M\rightarrow \infty$, it is equivalent to both our DFLVM and the LC approach, as $U_n \rightarrow U^*$ in this case.

For our simulations, we take the DFLVM model (Equations~\ref{eq:madermodel1} and \ref{eq:DFLVM}) as the true data generating process. Simulation parameters are presented in Table~\ref{tb:maderpars}.  
\begin{table}[H]
\label{tb:maderpars}
\caption{Simulation conditions. We use a full factorial design over the varying conditions for a total of $2^3 \times 3 = 24$ conditions. \vspace{0.1cm}}
\centering
\begin{tabular}{ccl}
\toprule
Condition & Value & Description \\
\midrule
\multicolumn{3}{l}{\textit{Varying conditions}} \\[2pt]
$M$           & 3, 10   & Measurements per participant \\
$\tau_{22}$   & 0.5, 1  & Standard deviation of $U_2$ \\
$\beta_2$     & 0, 0.6  & Slope for $U_2$ (normal model) \\
$\beta_2$     & 0, 0.25 & Slope for $U_2$ (Poisson \& binomial models) \\
$D'$ & Normal, Poisson, Binomial & Second-stage response distribution \\
\midrule
\multicolumn{2}{l}{\textit{Fixed conditions}} \\[2pt]
$S$         & 100 & Replicates per condition \\
$N$         & 500 & Number of participants \\
$\tau_{11}$ & 1   & Standard deviation of $U_1$ \\
$\alpha_1$  & 0   & Intercept for first-stage $\mu$ \\
$\alpha_2$  & 0   & Intercept for first-stage $\sigma$ (log-scale) \\
$\beta_0$   & 0   & Intercept for second-stage $\mu$ \\
$\beta_1$   & 0.6 & Slope for $U_1$ \\
$\rho$      & 0   & Correlation between $U_1$ and $U_2$ \\
$\xi$       & 1   & Second-stage error standard deviation (normal model) \\
\bottomrule
\end{tabular}
\end{table}

In all four competing estimation approaches, we use very weakly informative priors on the coefficients for the distributional covariates and keep the default priors provided in the \texttt{brms} package \parencite{burknerBrmsPackageBayesian2017} for the correlation matrix, intercepts, and scale parameters:
\begin{equation}
\begin{aligned}
\label{eq:maderpriors}
\mathbf{R} &\sim \text{LKJ}(1)\\
\beta_1,\,\beta_2 &\sim \text{Normal}(0,5)\\
\alpha_1,\,\alpha_2,\,\beta_0 &\sim \text{Student-t}(3, 0, 2.5)\\
\tau_{11},\,\tau_{22},\,\xi &\sim \text{Student-t}_{+}(3, 0, 2.5),
\end{aligned}
\end{equation}
where $\text{LKJ}(1)$ stands for the Lewandowski-Kurowicka-Joe distribution \parencite{lewandowskiGeneratingRandomCorrelation2009} with shape parameter 1, a value that provides a uniform prior over all appropriately sized correlation matrices. In the $2\times2$ case, this is equivalent to setting a $(0,1)$ uniform prior on $\rho$.

We first present the simulations results for the second-stage model with a normal likelihood and identity link, followed by the Poisson likelihood with log link. Results for the binomial likelihood with logit link can be found in Appendix~\ref{sec:appendix1}.
\begin{figure}[H]
\includegraphics[width=\textwidth]{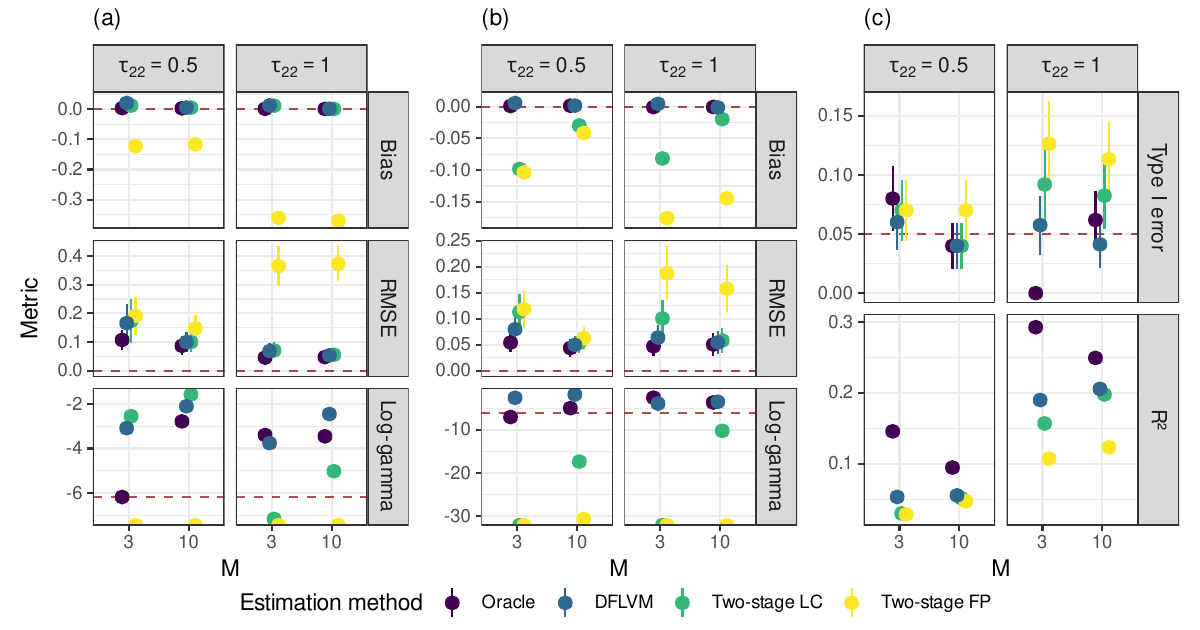}
\caption{Performance metrics for the second-stage normal likelihood model. (a) Estimation performance for coefficient $\beta_2$. (b) Same metrics for the standardized coefficient $b_2$. Dashed red lines indicate a reference value for the metric: 0 for bias and RMSE, 0.05 for type I error, and the critical value for the log-gamma statistic. For the latter, values under the line represent rejection of the null hypothesis of a well-calibrated posterior. (c) Model-level metrics in the second-stage normal likelihood model: Empirical type I error rate (nominal rate: 5\%) and predictive performance measured via out-of-sample $R^2$.}
\label{fig:sim1res1}
\end{figure}

\subsubsection*{Results: Normal second-stage response model}

Our results for parameter recovery and calibration (Figure~\ref{fig:sim1res1}a, b) show that the FP model generally underperforms across all conditions, regardless of which version of the estimate is considered. For the LC model, differences become apparent in terms of bias and calibration only when the standardized coefficient is considered. Because the DFLVM estimate has increased uncertainty, differences in RMSE are less stark and otherwise mirror differences in bias.
\begin{figure}[h]
\includegraphics[width=\textwidth]{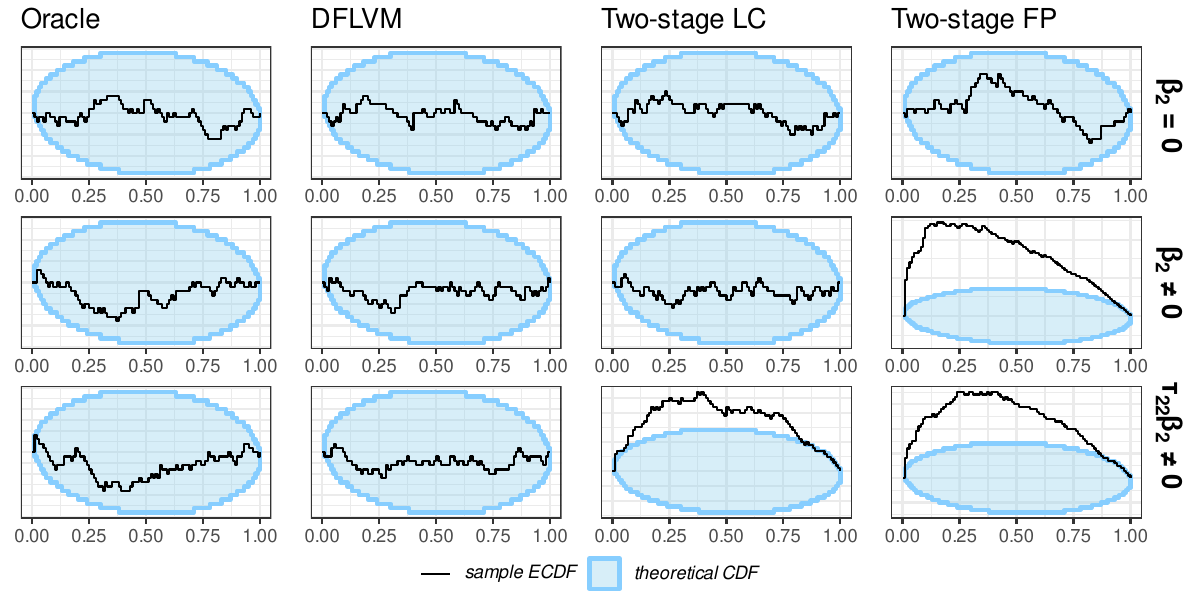}
\caption{ECDF difference plot for a selected simulation condition ($M = 10$, $\tau_{22}=0.5$) in the normal second-stage model. The blue areas represent the 95\%-confidence bands for a well-calibrated posterior.}
\label{fig:sim1sbc}
\end{figure}

Type I error and $R^2$ are shown separately in Figure~\ref{fig:sim1res1}c as these metrics do not depend on whether the coefficient is standardized. Differences between the models become more apparent with increasing random effects variance. The FP model again exhibits inferior performance. Although the LC model achieved similar performance to the DFLVM in parameter recovery of the unstandardized coefficient, it has slightly worse type I error and $R^2$.

We use the graphical approach of \cite{sailynojaGraphicalTestDiscrete2022} to further examine calibration of the resulting posteriors. Figure~\ref{fig:sim1sbc} shows that uncertainty is overall well-calibrated for every model in the null effect case ($\beta_2 = 0$). When the effect is not null, the sharp peak of the FP model indicates a clear underestimation of the overall effect size in both the unstandardized and standardized cases. For the LC model, a more subtle degree of underestimation appears only in the standardized version of the coefficient. No detectable differences appear between our approach and the oracle model, confirming that uncertainty is well-calibrated in both cases.

\begin{figure}[h]
\includegraphics[width=\textwidth]{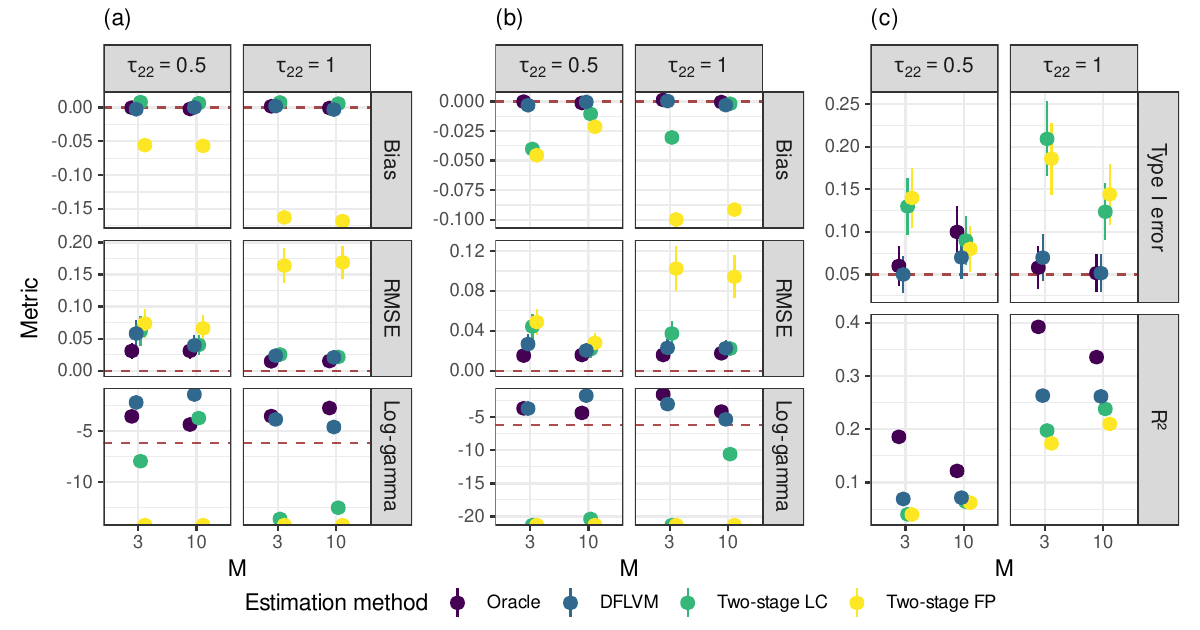}
\caption{Estimation performance in the second-stage Poisson likelihood model for coefficient $\beta_2$. The dashed red line indicates a reference value for each metric: 0 for bias and RMSE, 0.05 for type I error, and the critical value for the log-gamma statistic.}
\label{fig:sim1res1b}
\end{figure}

\begin{figure}[h]
\centering
\includegraphics[width=\textwidth]{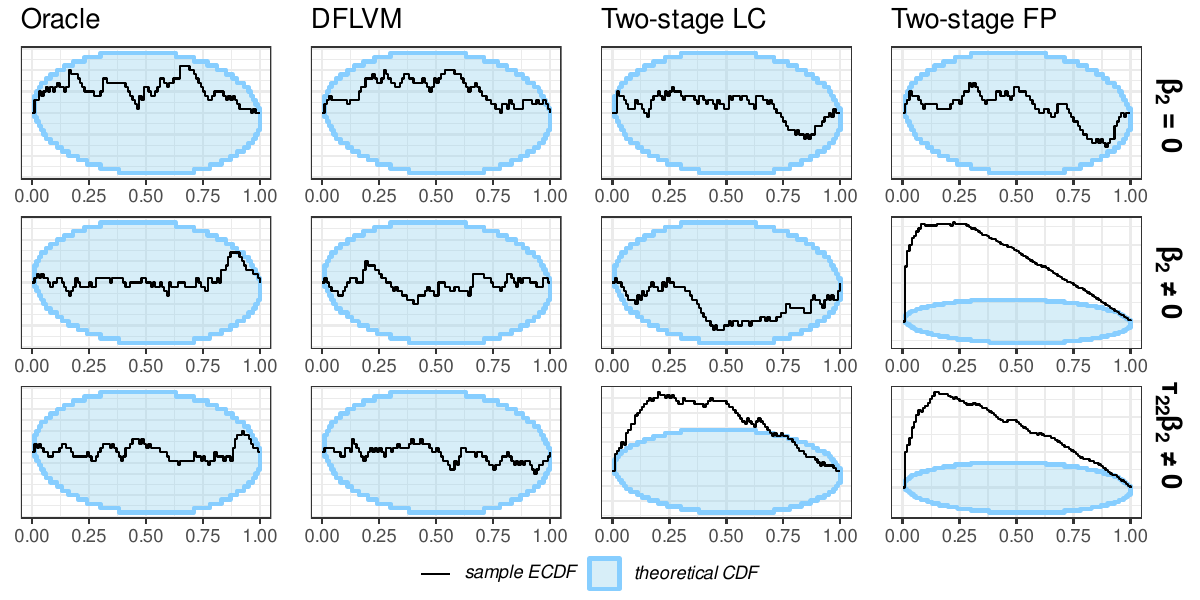}
\caption{ECDF difference plot for a selected simulation condition ($M = 10$, $\tau_{22}=0.5$) in the Poisson second-stage model. The blue areas represent the 95\%-confidence bands for a well-calibrated posterior.}
\label{fig:sim1sbcb}
\end{figure}

\subsubsection*{Results: Poisson second-stage response model}

For the Poisson model, the results in terms of parameter recovery, shown in panels a and b of Figure~\ref{fig:sim1res1b}, appear to be qualitatively similar to those of the normal model. Figure~\ref{fig:sim1res1b}c shows that inflation of the type I error rate is more pronounced for both two-step models. Predictive performance measured via $R^2$ again favors the DFLVM compared to the two-step approaches. The ECDF difference plot in Figure~\ref{fig:sim1sbcb} also reveals similar patterns of miscalibration as those seen in the normal model.

\begin{figure}[h]
\includegraphics[width=\textwidth]{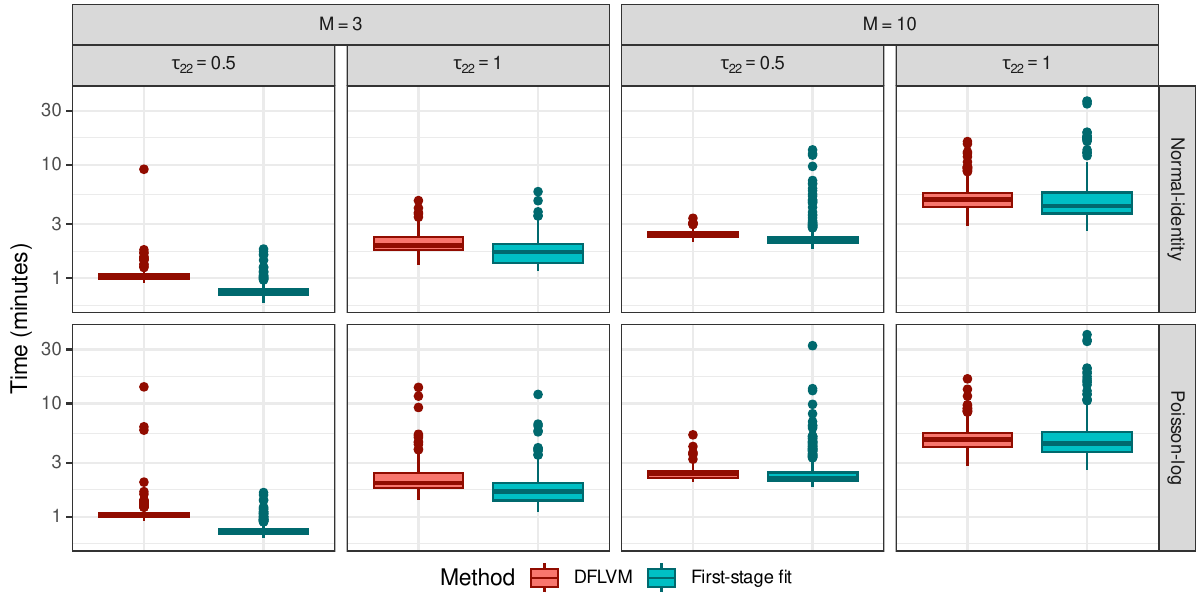}
\caption{Distribution of estimation times of the full DFLVM model against the first-stage fit shared by the two-step models. Time spent on the second stage of the two-step models is negligible and is therefore omitted here. The y-axis is log-scaled.}
\label{fig:sim1res4}
\end{figure}

\subsubsection*{Results: Estimation times}

Given the overall modest differences in performance between our DFLVM and the LC approach, one may be concerned that fitting the more complex model may incur unnecessary computational cost. Our results in Figure~\ref{fig:sim1res4} show that the median time to fit a DFLVM is only moderately higher than in the LC approach. In fact, the latter tends to have longer tails as the information in the model grows (i.e., higher random effects variance or more observations per person). Therefore, at least for the conditions assessed in this simulation study, the DFLVM appears to provide improved uncertainty quantification at an acceptable computational cost.


\section{Illustrative example}
\label{sec:casestudy} 

Having investigated the performance of the DFLVM approach under controlled conditions, we move on to investigating if practically meaningful differences also manifest in a realistic setting. We expand on the last analysis of \textcite{maderEmotionalInstabilityNeuroticism2023} by comparing the DFLVM to the authors' two-step LC approach on their \emph{Goettingen Ovulatory Cycle Diaries 2} (GOCD2) dataset.

A detailed description of the data collection process for the GOCD2 dataset is provided in \textcite{arslanRoutinelyRandomizePotential2021}. Briefly, the study recruited women who filled out a form at recruitment, which included a personality questionnaire, and who were subsequently asked to fill out an online diary for up to 70 days, from which a daily negative affect measurement could be constructed. The data, as made available in the replication package for \textcite{maderEmotionalInstabilityNeuroticism2023}\footnote{Found in the \texttt{Own diary data} folder at \url{https://osf.io/9dcxn/}, processed according to the code in \texttt{CodeFiles/OwnDiaryDataset13.Rmd}.}, contains 444 subjects after cleaning, with a median of 15 diary entries per participant (IQR: 15 -- 15, range: 4 -- 16).

Following the authors' specification, the first-stage model introduces censoring to account for floor and ceiling effects in the measurements of mood. The difference with the model used for our simulations in Equation~\ref{eq:madermodel1} is that the last line is replaced by
\begin{equation}
\begin{aligned}
    \text{Mood}^*_{nm} &\sim
    \text{Normal}(\mu_n,\,\sigma_n)\\
\text{Mood}_{nm} &=
    \begin{cases}
    1 & \text{Mood}^*_{nm} \leq 1\\
    \text{Mood}^*_{nm} & 1 < \text{Mood}^*_{nm} < 5\\
    5 & 5 \leq \text{Mood}^*_{nm},
    \end{cases}
\end{aligned}
\end{equation}
Where $\text{Mood}^*_{nm}$ represents the uncensored, unobservable response and $\text{Mood}_{nm}$ is the observed response, measured through an item scale with 5 options.

\begin{figure}[H]
\includegraphics[width=\textwidth]{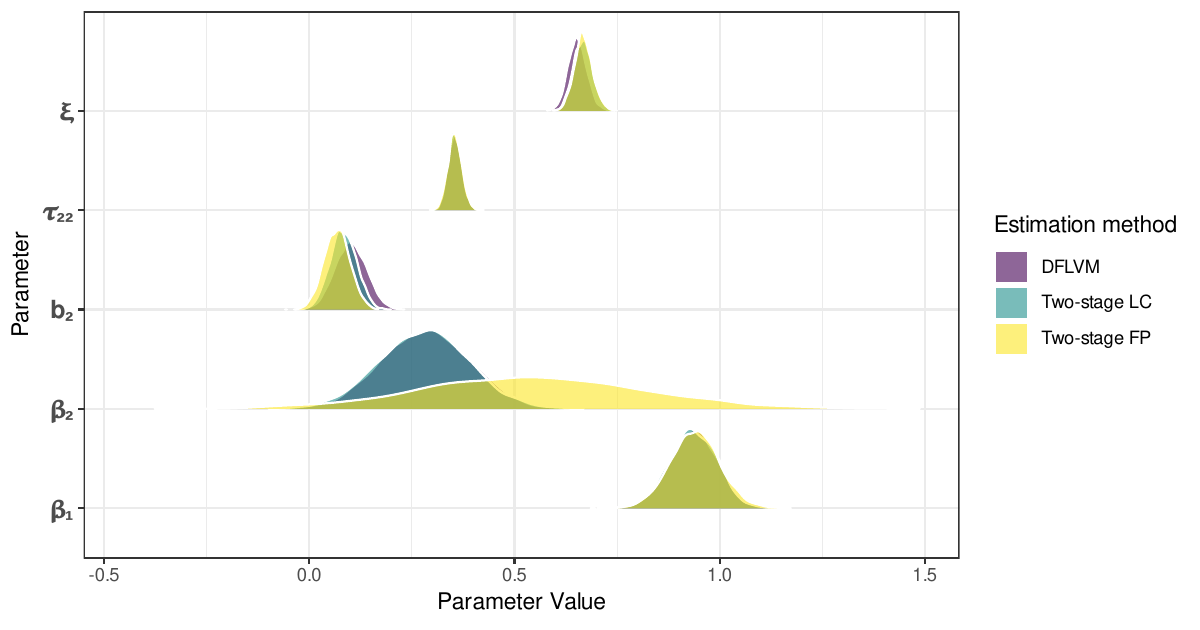}
\caption{Results for the reanalysis of \textcite{maderEmotionalInstabilityNeuroticism2023}. Density plots show the posterior distributions of key model parameters.}
\label{fig:res1fig1}
\end{figure}

We fit the authors' original two-step LC approach, our single-stage DFLVM and additionally investigate the fit of the two-stage Full Parameter (FP) approach. The procedure for standardization of coefficients was described in Section~\ref{sec:ddm-covariates}. Results for the key model parameters are shown in Figure~\ref{fig:res1fig1}. A difference in the unstandardized coefficient ($\beta_2$) for the FP model is expected, as the scale of the parameter is different from that of the random intercepts. However, when the standardized coefficient ($b_2$) is considered instead, all three methods produce a very similar estimate. Consistent with our simulations, a moderate downward bias is observed on the standardized coefficients for the two-stage approaches. Table~\ref{tb:maderest} shows the posterior estimates and in-sample $R^2$. The 95\% credible intervals exclude zero in all cases except for the unstandardized FP estimate. For $R^2$, we find the same ordering as in our simulations, showing that the DFLVM achieves better predictive performance, although the magnitude of the difference is small.
\begin{table}[!h]
\caption{First two rows show posterior summaries for the coefficient of the distributional covariate in unstandardized ($\beta_2$) and standardized ($b_2$) form. Last row shows in-sample point estimate of $R^2$. \vspace{0.1cm}}
\centering
\begin{tabular}[t]{llrr}
\toprule
Parameter & Model & Mean & 95\% CI\\
\midrule
\multirow{3}{*}{$\beta_2$} & DFLVM & 0.284 & {}[0.074, 0.495]\\
 & Two-step LC & 0.280 & {}[0.062, 0.499]\\
 & Two-step FP & 0.545 & {}[-0.002, 1.093]\\
\midrule
\multirow{3}{*}{$b_2$} & DFLVM & 0.101 & {}[0.026, 0.175]\\
 & Two-step LC & 0.083 & {}[0.021, 0.145]\\
 & Two-step FP & 0.068 & {}[0.006, 0.130]\\
\midrule
\multirow{3}{*}{$R^2$} & DFLVM & 0.402 & \\
 & Two-step LC & 0.392 & \\
 & Two-step FP & 0.390 & \\
\bottomrule
\end{tabular}
\label{tb:maderest}
\end{table}
The lack of any large differences between the DFLVM and the LC approaches is explained by the larger amount of per-person observations: participants had $15$ diary entries in the majority of cases, compared to $3$ and $10$ as evaluated during our simulations. 
Additionally, the error scale parameter for the second-stage response ($\xi$) has a posterior mean $0.65$, which is smaller than the value of $1$ we investigated. With regards to the FP approach, our simulations also showed that the difference is influenced by the random intercept scale $\tau_{22}$. For this dataset, $\tau_{22}$ has posterior mean of $0.36$, again smaller than our simulated values of $0.5$ and $1$. Taken together, these factors lead to reduced estimation error of the random intercepts and impact of the misspecified form of the distributional covariate.

As an additional investigation, we fit the same models in a subset of the original dataset, keeping only three diary entries per participant to match our simulation setup. The results are qualitatively similar and are presented in Appendix~\ref{sec:appendix2}.

\section{Discussion}
\label{sec:discussion}

We have introduced a framework for single-step estimation of models with distributional covariates. This addresses a long-standing gap in the psychology modeling toolkit, where interest in the association between person-level distributional features and other outcomes of interest have historically driven researchers to either use multi-step estimation procedures or ad hoc model implementations \parencite{ramExaminingInterindividualDifferences2005,jahngAnalysisAffectiveInstability2008,frischkornWorstPerformanceRule2016,maderEmotionalInstabilityNeuroticism2023}. Our framework generalizes past instances of models that were used to investigate associations with distributional features by introducing the notion of distributional covariates. That is, random intercepts that model inter-individual variation in distributional parameters, which can be simultaneously used to predict other downstream outcomes. By using a Bayesian framework, our model can be estimated in a single step and avoids issues pertaining to omission of estimation error / uncertainty propagation. We implemented our DFLVM framework in the \texttt{brms} package \parencite{burknerBrmsPackageBayesian2017}, a widely used interface for Bayesian model specification based on Stan \parencite{standevelopmentteamStanReferenceManual2026}. This means that our framework can be combined with other advanced modeling approaches available therein (e.g., missing data, censoring, custom distributions, among others), giving researchers a flexible and reliable platform on which to build their models.

In those cases where practitioners were able to implement single-step estimation approaches, they reported both two-step and single-step estimates for the data at hand \parencite{ramExaminingInterindividualDifferences2005, wangInvestigatingInterindividualDifferences2012}, but more extensive simulation studies to characterize the "region of concern" were not conducted. As a result, researchers remain wary of the estimates they obtain in such manner, leading to less confident claims about their results \parencite[e.g.,][]{maderEmotionalInstabilityNeuroticism2023}. Our simulation results clarify the contexts in which the omission of estimation error in two-step procedures can have a practical impact. We find that estimation error disappears quickly with the number of measurements per individual, more so when the measurements themselves have a low level of non-systematic error (i.e., high reliability). This is in line with previous empirical findings and the results of our own reanalysis of \textcite{maderEmotionalInstabilityNeuroticism2023}, which demonstrate that researchers often already collect enough per-person measurements for two-step procedures to produce accurate results.

On the other hand, one aspect that has seen less discussion in analyses that involve distributional covariates is the impact that the form of the covariate can have on model fit. Our DFLVM formulation highlights the fact that between-person variation can be regarded as a latent variable, different from the raw distributional summary it is associated with. More importantly, we show that misspecification of that form can have a larger impact on parameter recovery and model fit than estimation error. In our simulations, we chose to represent between-person variation as random intercepts in a log-scaled linear predictor for illustrative purposes. However, our reanalysis of \textcite{maderEmotionalInstabilityNeuroticism2023} showed that such a choice also appeared to be optimal in the context of a real dataset, as the FP approach obtained an inferior $R^2$ compared to the DFLVM and LC. This suggests that previous efforts which have investigated the associations of distributional summaries with other outcomes of interest \parencite[e.g.,][]{nielsenMeanCanWe2023}, may have failed to detect relevant patterns by not considering alternative forms of representing those summaries. 

A valid practical concern for the use of the DFLVM framework is the added computational cost of fitting a more complex model. Early comparisons found stark differences in the time it took to perform single-step compared to two-step estimation  \parencite[e.g., 30 hours vs 10 minutes in][]{ramExaminingInterindividualDifferences2005}, which could reasonably discourage practitioners. That said, ongoing developments in sampling techniques and the general increase of computing power continue to decrease estimation times. Our simulations demonstrate that single-step estimation takes a comparable amount of time relative to the first stage of a two-step procedure, with only a slight increase in median time. We chose to use Stan's MCMC sampler for posterior estimation due to its well-validated accuracy \parencite{pmlr-v258-magnusson25a}, but it could be relevant to investigate the potential of alternative approaches such as variational inference \parencite{bleiVariationalInferenceReview2017, zhang2022pathfinder} or simulation-based inference \parencite{cranmerFrontierSimulationbasedInference2020, zammit-mangionNeuralMethodsAmortized2024} to further reduce estimation times.

One limitation of the DFLVM framework as presented here is that it requires the construction of a likelihood that explicitly features the distributional covariates as parameters therein. In the case we have examined, the distributional feature of interest was variability of a normal response, and this made it straightforward to introduce random intercepts in the log-linear predictor of the scale parameter. For more general summaries, as the ones explored by \textcite{dejonckheereComplexAffectDynamics2019} and \textcite{nielsenMeanCanWe2023}, this task can be more challenging. Likelihood-free methods, such as those featured by simulation-based inference, can provide a helpful alternative in this case and should be explored in future work.

\section*{Acknowledgments}

This research was partially funded by Deutsche Forschungsgemeinschaft (DFG, German Research Foundation) Project 497785967. We gratefully acknowledge the computing time provided on the Linux HPC cluster at Technical University Dortmund (LiDO3), partially funded in the course of the Large-Scale Equipment Initiative by the DFG Project 271512359. We thank Nina Mader and Ruben Arslan for their guidance with the dataset and code we used for our illustrative example.

{\singlespacing
\printbibliography}
\newpage
\section*{Appendix}
\pagenumbering{roman} 
\renewcommand{\thesubsection}{\Alph{subsection}}
\renewcommand\thesection{\Roman{section}}
\renewcommand\thefigure{\thesubsection.\arabic{figure}} 
\setcounter{figure}{0}
\setcounter{table}{0}
\subsection{Simulation results for the second-stage binomial response model}\label{sec:appendix1}

\begin{figure}[H]
\includegraphics[width=\textwidth]{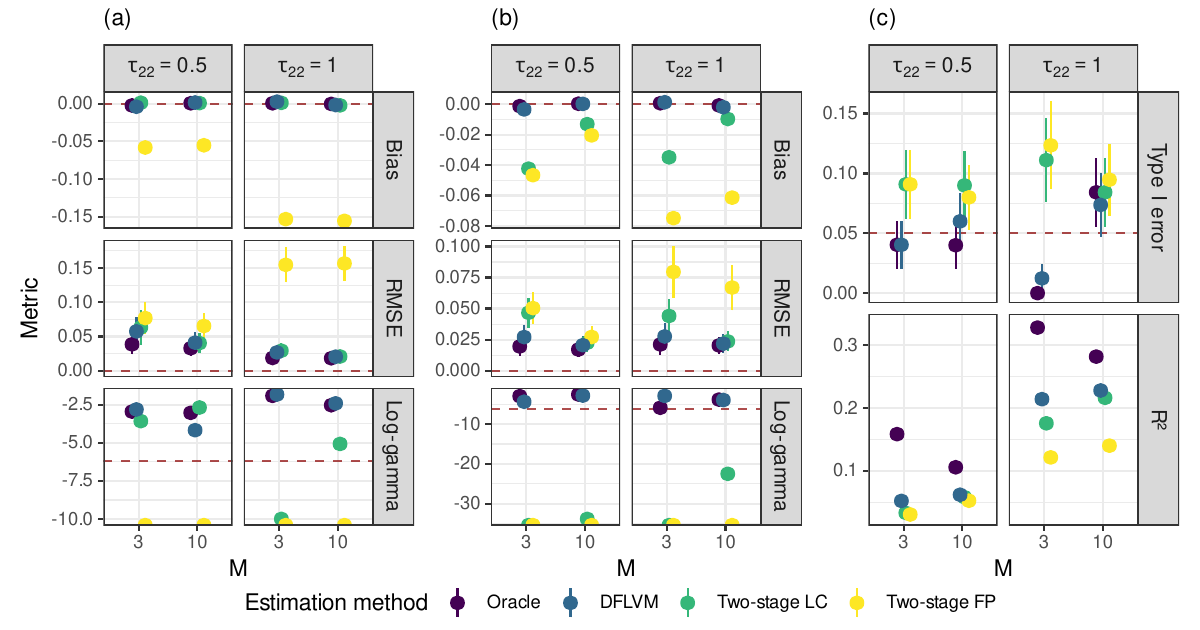}
\caption{Performance metrics for the second-stage binomial likelihood model. (a) Estimation performance for coefficient $\beta_2$. (b) Same metrics for the standardized coefficient. (c) Model-level metrics in the second-stage normal likelihood model: false positive rate and predictive performance measured via out-of-sample $R^2$.}
\label{fig:sim1A1}
\end{figure}

\begin{figure}[H]
\includegraphics[width=\textwidth]{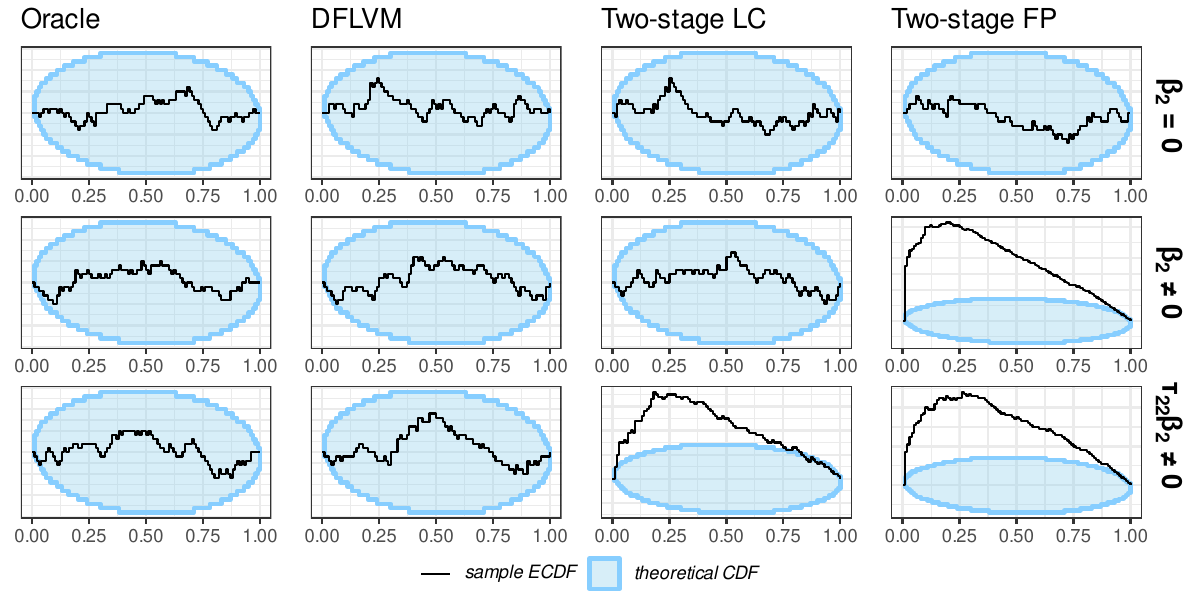}
\caption{ECDF difference plot for a selected simulation condition ($M = 10$, $\tau_{22}=0.5$) in the binomial second-stage model. The blue areas represent the 95\%-confidence bands for a well-calibrated posterior.}
\label{fig:sim1A2}
\end{figure}

\subsection{Results for reanalysis on subset of \cite{maderEmotionalInstabilityNeuroticism2023}}
\label{sec:appendix2}

These results were obtained by following the same procedures described in Section~\ref{sec:casestudy}, except that the dataset was filtered to retain only three diary measurements per person.

\begin{figure}[H]
\includegraphics[width=\textwidth]{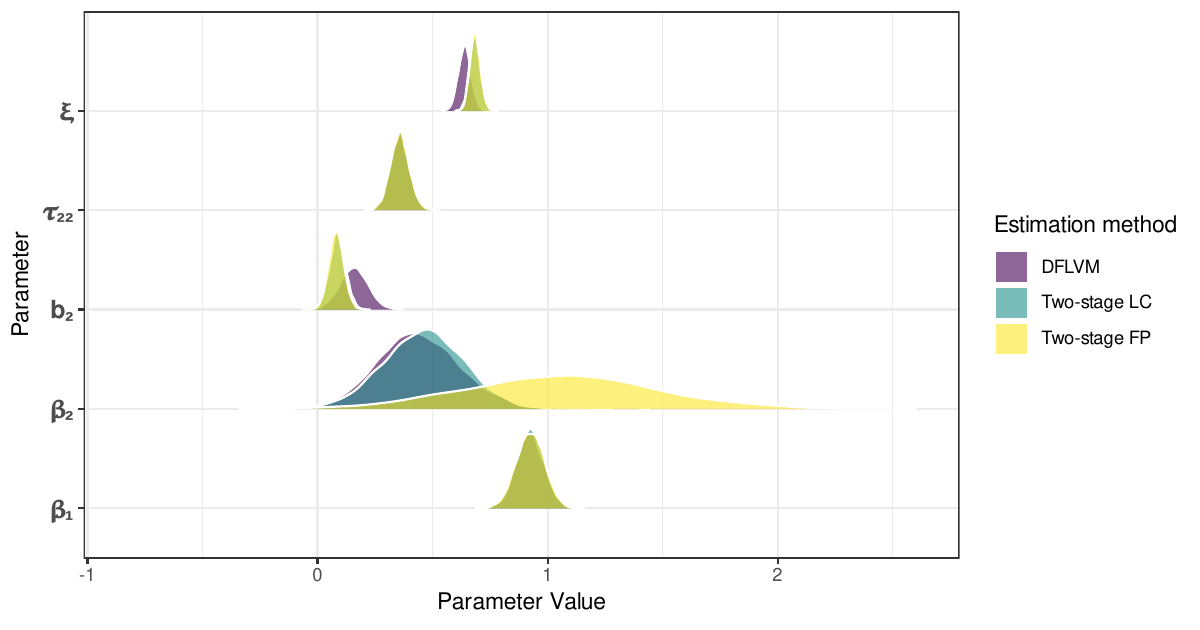}
\caption{Posterior distributions for key model parameters. }
\label{fig:res1fig1sub}
\end{figure}

\begin{table}[H]
\label{tb:maderestsub}
\caption{First two rows show posterior summaries for the coefficient of the distributional covariate in unstandardized ($\beta_2$) and standardized ($b_2$) form. Last row shows in-sample point estimate of $R^2$. \vspace{0.1cm}}
\centering
\begin{tabular}[t]{llrr}
\toprule
Parameter & Model & Mean & 95\% CI\\
\midrule
\multirow{3}{*}{$\beta_2$} & DFLVM & 0.446 & {}[0.096, 0.842]\\
 & Two-step LC & 0.460 & {}[0.111, 0.800]\\
 & Two-step FP & 1.060 & {}[0.250, 1.898]\\
\midrule
\multirow{3}{*}{$b_2$} & DFLVM & 0.157 & {}[0.034, 0.277]\\
 & Two-step LC & 0.086 & {}[0.018, 0.155]\\
 & Two-step FP & 0.082 & {}[0.017, 0.147]\\
\midrule
\multirow{3}{*}{$R^2$} & DFLVM & 0.370 & \\
 & Two-step LC & 0.359 & \\
 & Two-step FP & 0.359 & \\
\bottomrule
\end{tabular}
\end{table}
\end{document}